# NoxTrader: LSTM-Based Stock Return Momentum Prediction for Quantitative Trading

Wei-Ning, Chiu
Data *dept. NoxLab*
*Computer Science dept. NTHU*
Hsinchu, Taiwan
william08162010@gmail.com

Hsiang-Hui, Liu
Infrastructure *dept. NoxLab*
*Computer Science dept. NTHU*
Hsinchu, Taiwan
morrisliuting@gmail.com

Han-Jay, Shu
Data *dept. NoxLab*
*Electrical Engineering and Computer Science dept. NTHU*
Hsinchu, Taiwan
jerryshu98@gmail.com

**Abstract**— We introduce NoxTrader, a sophisticated system designed for portfolio construction and trading execution with the primary objective of achieving profitable outcomes in the stock market, specifically aiming to generate moderate to long-term profits. The underlying learning process of NoxTrader is rooted in the assimilation of valuable insights derived from historical trading data, particularly focusing on time-series analysis due to the nature of the dataset employed. In our approach, we utilize price and volume data of US stock market for feature engineering to generate effective features, including Return Momentum, Week Price Momentum, and Month Price Momentum. We choose the Long Short-Term Memory (LSTM) model to capture continuous price trends and implement dynamic model updates during the trading execution process, enabling the model to continuously adapt to the current market trends. Notably, we have developed a comprehensive trading backtesting system — NoxTrader, which allows us to manage portfolios based on predictive scores and utilize custom evaluation metrics to conduct a thorough assessment of our trading performance. Our rigorous feature engineering and careful selection of prediction targets enable us to generate prediction data with an impressive correlation range between 0.65 and 0.75. Finally, we monitor the dispersion of our prediction data and perform a comparative analysis against actual market data. Through the use of filtering techniques, we improved the initial -60% investment return to 325%.

## I. Introduction

In an era characterized by rapid advancements in algorithmic and machine learning-based trading, financial markets are undergoing a profound transformation. While many papers discuss how to enhance market prediction accuracy using artificial intelligence models, there is a scarcity of literature addressing their practical application in real markets or their failures when applied in such contexts. Therefore, this paper is dedicated not only to creating highly accurate models but also to achieving successful real-world applications in financial markets as the ultimate objective.

Certain existing papers provide us with a strong knowledge foundation and offer examples of integrating artificial intelligence models with financial market data. "Machine Learning Approaches in Stock Price Prediction: A Systematic Review"[P. Soni, Y. Tewari and D. Krishnan , 2022][1] provides a profound overview of the current use of artificial intelligence models in market prediction, including machine learning models such as SVM and random forest, as well as deep learning models like LSTM and RNN, giving us an initial comprehensive understanding. Additionally, "101 Formulaic Alphas"[Z. Kakushadze, 2015][2] demonstrates methods of utilizing market data and emphasizes the significance of feature engineering in the field of financial market prediction.

Unlike prevailing methodologies, NoxTrader leverages the predictive capabilities of LSTM networks and supervised learning techniques to discern intricate patterns within historical data, effectively capturing the nuanced fluctuations in market prices. This novel target of prediction distinguishes us significantly from other research papers and entities employing LSTM models for market forecasting, resulting in a substantial enhancement in our market performance. NoxTrader introduces a unique perspective on label generation by adopting the concept of "return momentum", which is the difference of return between two consecutive days, as a predictive label with appropriate filter, as opposed to the more customary use of raw returns. This nuanced approach to label formulation adds an additional layer of sophistication to NoxTrader's predictive capabilities.

The subsequent sections of this paper are carefully structured to provide a comprehensive exposition of NoxTrader's inner workings. We delve into the intricacies of each constituent unit, meticulously detailing the process of feature generation, predictive modeling, and the creation of a robust backtest environment. Our narrative is further enriched by a diverse array of empirical experiments, designed to showcase the results garnered from NoxTrader's operational deployment and substantiate its potential profitability. Finally, we make a complete discussion and conclude it with some possible future improvements we're going to make.

## II. Methods

NoxTrader's implementation consists of three primary components. The initial component, termed "Feature Generation," assumes the role of crafting distinct facets of features and subsequently conveying them to the model. The subsequent component, denoted as the "Prediction" module, bears the responsibility of harnessing the previously generated features. It assigns scores to individual stocks within our screener based on these features. Following the evaluation of all stocks, a mechanism for portfolio construction and strategy backtesting becomes imperative. This function, referred to as "Backtest," facilitates the process of portfolio assembly and strategy evaluation. In the subsequent sections, a comprehensive exposition of each of these three components is presented.

## A. Feature Generation

The historical data supplied by the yfinance platform embodies a set of fundamental raw metrics, encompassing the opening, closing, highest, and lowest values of each individual stocks. Notably, these initial data points, while serving as a foundational basis, lack the requisite depth to facilitate effective learning and precise predictive outcomes within the model. As such, a compelling imperative exists to undertake a transformative process on this raw dataset, thereby bringing in some enlightening features that are primed to empower the model's learning capacity.

In the spirit of holistic pattern recognition, NoxTrader undertakes a multifaceted approach to feature engineering. We collect market data for US companies with a market capitalization of more than 200 billion dollars and draw upon a rich spectrum of insights that collectively enhance the model's ability. The subsequent enumeration explain the diverse array of features meticulously integrated into the NoxTrader framework:

- Returns: showing how much the closing price has grown compared to its previous value, usually from the day before. This measure helps the model understand the changes in stock values over time more effectively.

- ReturnMomentum: Building upon the Returns paradigm, ReturnMomentum augments the model's insight by quantifying the differential between the current day's Returns and the Returns observed on the preceding day. This parameter encapsulates the intraday dynamics that could potentially exert influence on future stock performance.

- ReturnAcceleration: Delving deeper into the temporal dynamics, ReturnAcceleration discerns the variance in ReturnMomentum from one day to the next, thereby encapsulating the intricate curvature of the stock valuation trajectory. This higher-order derivative augments the model's ability to capture evolving trends.

- WeekPriceMomentum: Harnessing a broader temporal horizon, WeekPriceMomentum appraises the growth rate of the closing price compared to its state a week prior. This temporal frame of reference imparts a long-range perspective on valuation trends, arming the model with insights into sustained momentum. [2]

- MonthPriceMomentum: Extending the purview even further, MonthPriceMomentum represents the growth rate of the closing price in relation to its value a month preceding. This elongated temporal context serves as a harbinger of extended trends, enriching the model's predictive prowess. [2]

- VolumeVelocity: Recognizing the pivotal role of trading volume, VolumeVelocity indexes the growth rate of trading volume in comparison to the preceding day. This feature offers insights into market sentiment and potential shifts in supply-demand dynamics. [3]

These intricate features collaboratively contribute to the enhancement of NoxTrader's learning efficacy. By seamlessly integrating these multifaceted metrics, the model is able to glean a more profound comprehension of market nuances, thereby enhancing its predictive power and heightening the precision of its projections.

## B. Prediction Method

In this section, we detail the methodology employed for predicting stock price changes using a Long Short-Term Memory (LSTM) network. The dataset, feature extraction, model architecture, loss function, performance evaluation, and the rationale behind the chosen approach are discussed.

- Dataset: The dataset comprises individual data instances, each consisting of two main components: features and labels. The features encapsulate information from the stock market for the past 10 days, including features mentioned in part A. Notably, the features incorporate data from the current day as well as the preceding 9 days. The labels represent the difference between return of two consecutive days, namely return momentum. A training set of 240 such instances is constructed, chosen based on utilizing historical data from the past year. Since stock market only opens 5 days a week, approximately 20 trading days correspond to a month.

- Model Architecture: Our approach employs the Long Short-Term Memory (LSTM) network as the primary model architecture. The LSTM's inherent ability to capture temporal dependencies makes it suitable for modeling stock price patterns. By recognizing the analogy between stock prices and language, which both exhibit temporal sequences, the LSTM aims to capture intricate patterns in stock price fluctuations.

- Loss Function: The Mean Squared Error (MSE) is chosen as the loss function for the LSTM model. This selection aligns with the objective of minimizing the discrepancy between predicted and actual stock return momentum. The MSE quantifies the average squared difference between predicted and actual values, enabling the model to learn optimal parameters that minimize this error.

- Prediction Performance Evaluation: To assess the performance of the model, we not only employ the MSE loss function but also calculate the correlation between our predictions and the true labels. This correlation metric ranges predominantly between 0.65 and 0.75, signifying a meaningful correspondence between predicted and actual trends. Importantly, considering the temporal nature of stock data, retraining the model is necessary every 10 days to ensure its adaptability to evolving market patterns.

- Model Generalization: While a predictive horizon beyond 10 days might be appealing, we observed a decline in correlation beyond this point. Specifically, if we use the same model to make predictions for 40 days, the correlation for the initial 20 days significantly surpasses the latter 20 days, implying reduced accuracy for longer prediction horizons. Consequently, forecasting stock prices over an entire year would necessitate the training of 24 separate models, each specialized for a specific 10-day prediction window.

## C. BackTest Environment

The presented backtesting framework is designed with the intention of harnessing the outcomes generated by our model. This is accomplished through the conversion of model-generated labels into corresponding stock positions, followed by the simulation of trading activities under conditions resembling those of the real market. The outlined system comprises two sections: label-to-position conversion and performance evaluation.

i. Label-to-position Conversion: This section consists of a two-fold procedure, involving filtration and capital allocation.

- Filtration: Stocks whose predicted labels fall within a predetermined range are classified as "no-trade" due to our observation of heightened correlation between predicted and actual labels when the predicted label magnitudes are large. Additionally, it has come to our attention that a strong concordance exists between the polarity of returns velocity and returns, particularly evident in scenarios where the magnitude of returns velocity are large. Following the identification of "no-trade" entities, the remaining candidates will undergo an adjustment process, involving a subtraction by a constant if the value is positive, and an addition by a constant if the value is negative.

- Capital Allocation: The remaining candidates will partake in a weighted average computation. On a daily basis, each candidate will be allocated a position value. The position value is determined by multiplying the total equity by their respective weighted average.

ii. Performance Evaluation: The computation of portfolio gains and losses entails the multiplication of positions with actual market returns. The subsequent benchmarks are introduced to provide enhanced comprehension of the portfolio's performance.

- In the market days: The ratio of the number of days in the market to the total number of days.

- Position Qualified: The ratio of the number of position in the market to the total number of position.

- Annual Returns: the total returns earned by an investment in a year, considering compounding effects.

- Win Rate: the percentage of successful trades among all position.

- Max Drawdown: the largest percentage decline in an investment's value from its peak to the lowest point.

## III. EXPERIMENT RESULT

The following experiment will demonstrate our approach to selecting model labels, establishing evaluation criteria, and ultimately formulating a comprehensive strategy.

### A. Initial Returns Prediction

During the first stage of our research, a noteworthy discovery was made. Despite our efforts, the correlation between predicted results (returns) and actual outcomes had become nearly negligible. This puzzling development extended to post-hoc testing, where the model's performance remained far from satisfactory. Although the overall correlation stood at around 0, it was intriguing to observe that the predicted trends of upward and downward movements exhibited some semblance to the true labels. To probe deeper, an analysis was conducted by computing the correlation between the predicted return differential and corresponding true label differentials, yielding an improved but modest correlation coefficient of 0.2. Please refer to Figure 1. The solid line represents the return momentum of true market while the broken line represents the differential of predicted return.

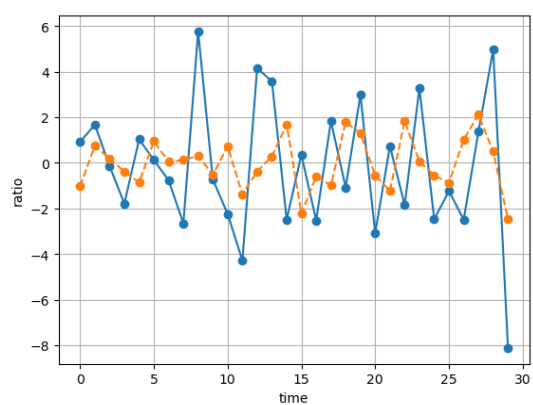

Fig. 1. Correlation between the Predicted Returns Differential and the Corresponding True Return Momentum

### B. Refining Label Representation

Upon entering this stage, a pivotal change was introduced in the way we represented the true labels. Shifting from the initial approach of using stock returns, we adopted return momentum as the new label representation. This alteration yielded remarkable results as the correlation coefficient surged to an impressive 0.6. Please refer to Figure 2. The solid line represents the return momentum of true market while the broken line represents the predicted return momentum. This change reaffirmed the importance of label representation in the predictive accuracy of the model. However, a concerning pattern emerged when the predicted results were inversely transformed to the original returns. The correlation plummeted back to negligible levels, puzzling us further.

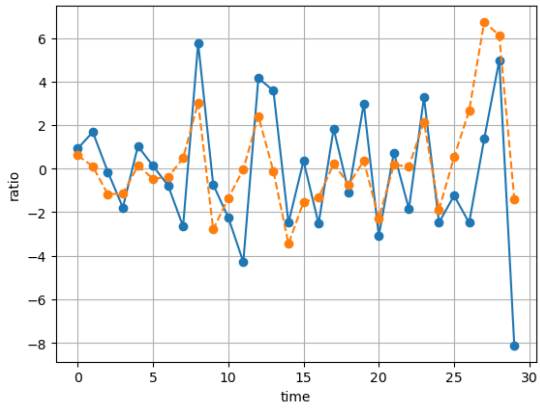

Fig. 2. Correlation between Predicted Return Momentum and True Labels

*C. Return Momentum Backtesting*

Our focus shifted towards understanding the stack discrepancy observed during the transformation from differentials to actual price changes. Intriguingly, employing the true return momentum as inputs for backtesting proved highly effective. Please refer to Figure 3. This unexpected success hinted at an inherent capability of return momentum to profit effectively without transforming back to return. Despite this, using the predicted return momentum for backtesting yielded disappointing results, indicating a major discrepancy between the model's predictive power and its application to the actual data. Please refer to Figure 4.

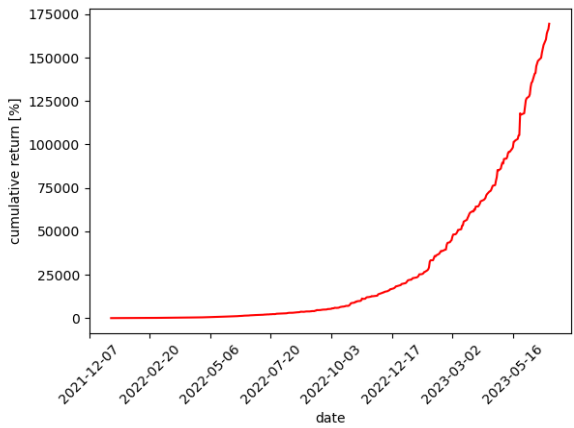

Fig. 3. True Label of Return Momentum as Inputs for Backtesting

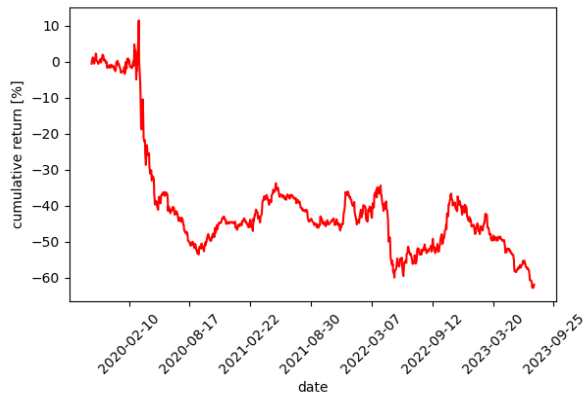

Fig. 4. Predicted Return Momentum as Inputs for Backtesting

*D. Correlation Viability Assessment*

To evaluate the viability of the correlation between predicted and true labels, stocks exhibiting a correlation exceeding 0.7 within each group are selected for a backtesting exercise. Each group represents a four-month testing period. The findings reveal that the observed correlation is indeed viable, substantiated by an annual return of 122.95% and a maximum drawdown of 11.14%. The comprehensive results of this analysis are illustrated through Figure 5.

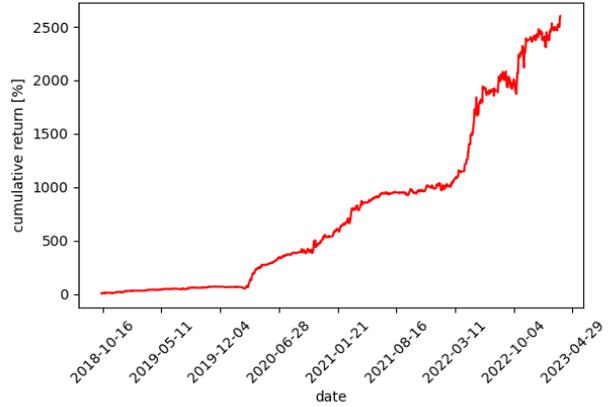

Fig. 5. Cumulative Return Chart of Candidates with High Correlation

*E. Correlation-Label Relationship*

Our investigation delves into establishing the connection between labels and correlation, building on the understanding that high correlation signifies superior performance. During this analysis, we incorporate the standard deviation of labels within each distinct group. It reveal a discernible pattern: instances of heightened standard deviation correlate with relatively high correlation. For an in-depth visual representation, please refer to Figure 6.

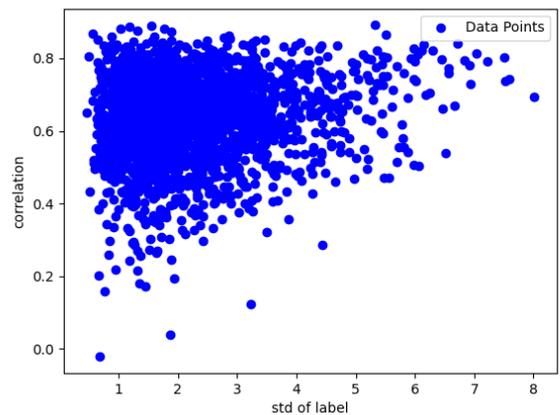

Fig. 6. Relation between Standard Deviation of Labels and Correlation

*F. Final Strategy*

The investigation into the Correlation-Label Relationship has bestowed insights that guide our strategic decisions. Labels characterized by little absolute values have been deemed less conducive to our strategy therefore discarded in the strategy. Furthermore, the remaining labels are subjected to an adjustment process, aligning their values closer to zero in a standardizing approach. The empirical outcomes are striking: a cumulative return of 325.38% over a span of six years, an annual return of 37.72%, and a maximum drawdown of 23.84%. The visual representation of a gracefully undulating curve, akin to an exponential curve, serves as compelling evidence of the pronounced positive impact of our filtration approach. A comprehensive tabulation can be found in Table I, while Figure 7 visually reinforces these outcomes.

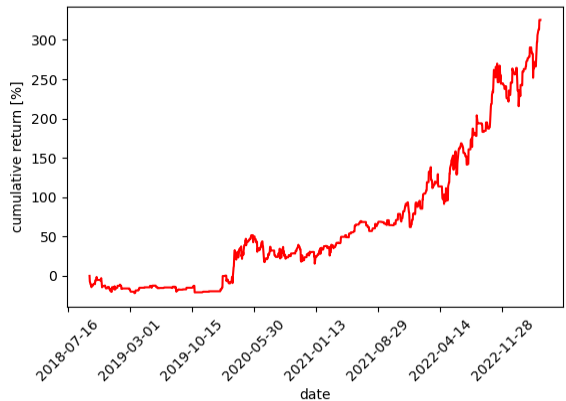

Fig. 7. Cumulative Return Chart of Final Strategy

TABLE I. RESULT DATA OF FINAL STRATEGY

| Property | Value |
| --- | --- |
| Start Date | 2018-10-04 |
| End Date | 2023-04-17 |
| Market Days | 1140 |
| In the Market Days | 41.491228 % |
| Position Qualified | 2.353445 % |
| Commision | 0.01 % |
| Equity Initial | 1000000 |
| Equity Final | 4253800 |
| Return | 325.380000 % |
| Ann. Return | 37.718672 % |
| Win Rate | 61.090909 % |
| Max. Drawdown | 23.837605 % |

## IV. CONCLUSION

With a paramount focus on stock market trading, NoxTrader's core objective centers on the cultivation of sustained moderate to long-term profits. Rooted in an intricate learning process, NoxTrader derives its insights from historical trading data through a steadfast reliance on time-series analysis, harmonizing seamlessly with the intrinsic characteristics of the employed dataset. In contrast to conventional methodologies, NoxTrader diverges by introducing an innovative approach to label generation. This entails the incorporation of the "return momentum" concept as a predictive label combined with carefully defined filters. This distinctive strategy yields notable outcomes that underscore its significance and efficacy.

In envisioning the future, there lie exciting avenues for further refinement and expansion. The pursuit of more robust predictive metrics, such as incorporating measures like standard deviation, promises to enhance the model's correlation with true labels, bolstering the effectiveness of its predictions. Moreover, further feature engineering will be conducted, with the aim of identifying features highly correlated with return momentum to enhance the predictive accuracy of the model. Additionally, we intend to apply the same methodology to not only the U.S. stock market but also to cryptocurrency and the Chinese stock market, which will serve as our next testing subjects. Ultimately, we aspire to implement this strategy model in the real market and conduct further observations.

In closing, NoxTrader stands as a testament to the potential within algorithmic trading models. Through this comprehensive exposition, it is evident that NoxTrader's journey is just beginning, with the horizon teeming with opportunities for innovation, refinement, and ever-greater achievements in the realm of model trading.